\documentclass[sigconf]{acmart}

\AtBeginDocument{%
  \providecommand\BibTeX{{%
    \normalfont B\kern-0.5em{\scshape i\kern-0.25em b}\kern-0.8em\TeX}}}

\setcopyright{none}
\copyrightyear{2018}
\acmYear{2018}
\acmDOI{XXXXXXX.XXXXXXX}

\usepackage{caption,nccmath}
\usepackage{subcaption}
\usepackage{subcaption}
\usepackage[titlenumbered,ruled]{algorithm2e}
\usepackage{multirow}
\usepackage{arydshln}
\usepackage{amsmath}

\usepackage{cleveref}
\usepackage{balance} 

\begin{document}

\title{Retentive Decision Transformer with Adaptive Masking for Reinforcement Learning based Recommendation Systems}

\author{Siyu Wang}
\affiliation{%
  \institution{The University of New South Wales}
  \city{Sydney}
  \country{Australia}
  \postcode{2052}
}
\orcid{0009-0008-8726-5277}
\email{siyu.wang5@student.unsw.edu.au}

\author{Xiaocong Chen}
\affiliation{%
  \institution{Data61, CSIRO}
  \city{Eveleigh}
  \country{Australia}
}
\email{xiaocong.chen@data61.csiro.au}

\author{Lina Yao}
\affiliation{%
  \institution{Data61, CSIRO}
  \city{Eveleigh}
  \country{Australia}
}
\affiliation{%
  \institution{The University of New South Wales}
  \city{Sydney}
  \country{Australia}
}
\email{lina.yao@unsw.edu.au}

\settopmatter{printacmref=false}

\renewcommand{\shortauthors}{Siyu Wang, Xiaocong Chen \& Lina Yao}

\begin{abstract}
Reinforcement Learning-based Recommender Systems (RLRS) have shown promise across a spectrum of applications, from e-commerce platforms to streaming services. Yet, they grapple with challenges, notably in crafting reward functions and harnessing large pre-existing datasets within the RL framework. Recent advancements in offline RLRS provide a solution for how to address these two challenges. However, existing methods mainly rely on the transformer architecture, which, as sequence lengths increase, can introduce challenges associated with computational resources and training costs. Additionally, the prevalent methods employ fixed-length input trajectories, restricting their capacity to capture evolving user preferences.
In this study, we introduce a new offline RLRS method to deal with the above problems. We reinterpret the RLRS challenge by modeling sequential decision-making as an inference task, leveraging adaptive masking configurations. This adaptive approach selectively masks input tokens, transforming the recommendation task into an inference challenge based on varying token subsets, thereby enhancing the agent's ability to infer across diverse trajectory lengths. Furthermore, we incorporate a multi-scale segmented retention mechanism that facilitates efficient modeling of long sequences, significantly enhancing computational efficiency. Our experimental analysis, conducted on both online simulator and offline datasets, clearly demonstrates the advantages of our proposed method.
\end{abstract}






\maketitle

\section{Introduction}
Reinforcement Learning (RL)-based Recommender Systems (RS) have emerged as powerful tools across diverse applications, from e-commerce and advertising to streaming services. Their strength lies in their ability to adapt to the dynamic nature of user interests in real-world scenarios~\cite{chen2021survey}. In RLRS, agents interact with environments, recommending items and receiving feedback in the form of rewards. Over time, these agents refine their policies to maximize long-term rewards, such as enhancing user satisfaction or engagement. Recently, \citet{chen2023opportunities} suggest that the offline RLRS would be a better solution than RLRS. The offline RLRS empowers the RL agent to learn from the pre-collected datasets instead of learning from interaction. With the offline RLRS, numerous datasets can be used to train the RL agent, which can significantly improve the training efficiency and performance of RLRS.
As a typical example, \citet{Wang2023} propose CDT4Rec that incorporates the offline RL into RS.
Although this method demonstrated potential, subsequent ablation studies underscored the substantial impact that input trajectory length—referred to as context length—has on the model's performance.


In the realm of RS, a user's trajectory comprises a chronological sequence of their interactions and behaviors, offering valuable insight into how their preferences evolve over time. This trajectory plays a pivotal role in capturing the changing trends and shifts in what a user might find appealing or engaging.
However, given the dynamic nature of user interests in RS, it becomes clear that the historical significance of these trajectories varies among users. For some, recent interactions may be the most indicative of their current interests, while for others, a longer history may provide a clearer view of their enduring preferences.
Considering this variability, there is a pressing need for RS models capable of intelligently and adaptively handling trajectories of different lengths. We address this problem by designing a novel adaptive causal masking mechanism. This adaptive capability is vital as it empowers the recommendation system to seamlessly switch between using recent interactions (for short-term insights) and incorporating broader historical patterns (for long-term insights). By doing so, the model can make more informed and nuanced recommendations, thereby enhancing its predictive accuracy and the relevance of its suggestions.

Another challenge of employing transformer-based offline RL in Recommender Systems, is the inherent complexity of Transformers. This complexity tends to escalate with increasing sequence lengths, leading to significant challenges in terms of memory usage, latency, and training expenses~\cite{shazeer2019fast, peng2023rwkv, sun2023retentive,agarwal2023transformers}.  Such challenges render them less practical for deployment in real-world scenarios, particularly within large-scale systems.
To address this, we draw inspiration from RetNet~\cite{sun2023retentive}, which facilitates efficient long-sequence modeling at a reduced inference cost. Building upon this, we introduce a novel framework: the Retentive Decision Transformer with Adaptive Masking for Offline Reinforcement Learning in Recommender Systems (MaskRDT). 


\begin{figure*}[h]
  \centering
  \includegraphics[width=\linewidth]{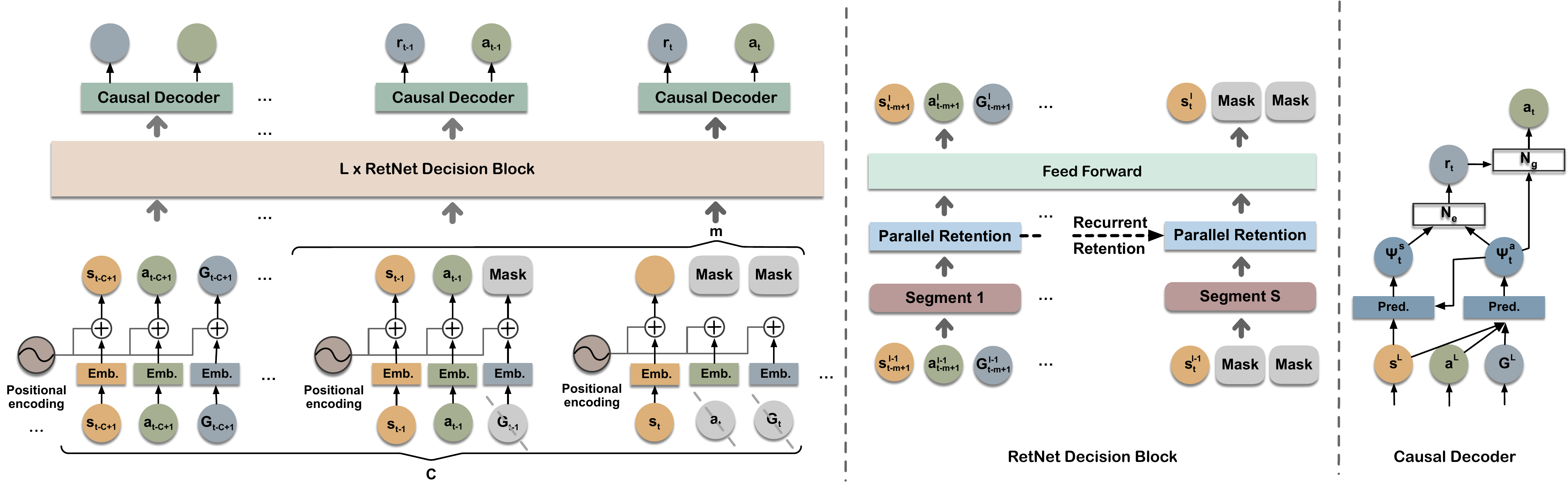}
  \caption{The comprehensive MaskRDT architecture. Starting from the left, states, actions and RTGs are transformed through linear embeddings, with an added absolute positional embedding. This trajectory segmentation, post-masking based on a predefined configuration, is fed into the initial retention block. The middle of the figure is the retention mechanism, where the masked trajectory is partitioned into \(S\) sub-segments. Computations within each segment are parallel, while recurrent retention computations bridge the segments. On the right, the causal layer emerges post the \(L\)-th block, producing two distinct representations directed into separate prediction layers. Crowning the architecture are two networks: \(N_e\) for reward estimation and \(N_g\) for action prediction.}
  \Description{.}
  \label{overrall}
\end{figure*}
In our MaskRDT framework, we reframe the RLRS challenge by treating sequential decision-making as an inference task with specific masking configurations. By strategically masking specific tokens within input trajectories, we dictate which portions of the user's history are visible to the model and which predictions it should generate. This method ensures that the model receives selective information about a user's past interactions and behaviors, guiding it to predict items for recommendation. A cornerstone of our approach is the innovative Adaptive Causal Masking. This technique introduces variability in the lengths of trajectories fed to the model, exposing the agent to a diverse range of trajectory segments. 
This adaptive strategy alternates between longer and shorter sequences, providing the agent with a nuanced and dynamic view of user behaviors. 
To bolster the efficiency of our model, we've integrated a multi-scale segmented retention mechanism, serving as an adept alternative to conventional multi-head attention. This design choice ensures long-sequence modeling is not only efficient but also resource-conscious. It achieves this by encoding each segmentation in parallel for swift computation, while different segmentations are encoded recurrently to reduce the training cost.

The key contributions of this study are as follows:
\begin{itemize}
    \item We model the offline RLRS challenge as an inference task using a unique masking configuration.
    \item Our innovative adaptive causal masking configuration allows the model to handle variable token lengths during training, enhancing its inference capabilities across diverse time frames.
    \item By integrating the causal retention network with masking, we achieve efficient long-sequence modeling while minimizing training costs.
    \item We empirically demonstrate the effectiveness of MaskRDT through comprehensive experiments on various datasets and in an online simulator.
\end{itemize}

\section{Problem Formulation}
\label{Formulation}
Given a set of users $\mathcal{U} = {u_0, u_1,...,u_n}$, a set of items $\mathcal{I} = {i_0, i_1,...,i_m}$, and historical interaction trajectories of users over a sequence of time steps $t = 1,...,T$, the goal of an RL-based Recommender System is to leverage historical interaction data to learn an effective policy $\pi$ that recommends items to users in a way that maximizes their satisfaction and overall engagement.

By transforming this scenario into an offline RL framework, we can employ the Markov Decision Process (MDP) paradigm~\cite{sutton2018reinforcement}. The core components of this MDP, tailored for recommendation, are:
\begin{itemize}
    \item State Space ($\mathcal{S}$): The state space represents the information and historical interaction context of users. In time step $t$, the state $s_t \in \mathcal{S}$ captures user features, preferences, and their historical interactions.
    \item Action Space ($\mathcal{A}$): The action space represents the choices available to the recommender agent in each state. $\mathcal{A}(s_t)$ denotes the set of actions possible in state $s_t$, where an action $a_t$ corresponds to recommending an item to a user.
    \item Transition Probability ($\mathcal{P}$): The transition probability \\
    $p(s_{t+1}|s_t, a_t) \in \mathcal{P}$ defines the likelihood of transitioning from state $s_t$ to $s_{t+1}$ when action $a_t$ is taken.
    \item Reward Function ($\mathcal{R}$): The reward function $\mathcal{R}(s, a) \rightarrow \mathbb{R}$ quantifies the immediate benefit of taking action $a$ in state $s$. In the context of the recommender system, the reward $r_t$ is based on the feedback received from users on recommended items.
    \item Discount Factor ($\gamma$): The discount factor $\gamma \in [0, 1]$ determines the weight of future rewards compared to immediate rewards in the agent's decision-making process.
\end{itemize}
An RL agent's objective is to learn a policy $\pi$, which is a mapping from states to actions, that maximizes the expected cumulative reward over trajectories:
\begin{equation}
J(\pi) = \mathbb{E}_{\tau \sim p_\pi(\tau)} \bigg[\sum_{k=0}^{\infty} \gamma^k r(s_t, a_t)\bigg],
\end{equation}
where $\tau = (s_0, a_0, s_1, a_1, ..., s_T, a_T)$ represents a trajectory under policy $\pi$.

In Offline RL, the focus is on improving the agent's policy using only a static dataset $\mathcal{D}$ of historical transitions, without further online interaction. The dataset $\mathcal{D}$ contains tuples $(s_t^u, a_t^u, s_{t+1}^u, r_t^u)$, where each tuple corresponds to a user $u$ at time step $t$. 
Formally, the problem can be summarized as learning a policy $\pi$ that maximizes the expected cumulative reward using the provided dataset $\mathcal{D}$, in which each transition tuple is sampled according to the distribution $\pi_{\beta}$.
By solving this Offline RL-based Recommender System problem, the aim is to make accurate and effective recommendations to users based on historical interactions, without the need for real-time exploration and interaction with the environment.

\section{Methodology}
\subsection{Structuring Trajectories in RL as Sequences}

RL-based RS traditionally interprets trajectories as sequences of state-action tuples as described in~\cref{Formulation}, capturing an agent's decisions and the associated outcomes.
However, following the perspective presented in~\cite{NEURIPS2021_7f489f64, janner2021offline}, we recast these trajectories as sequences of tokens. This refined representation emphasizes three critical tokens: states, actions, and returns-to-go (RTG), with the following trajectory representation:
\begin{equation}
\label{eq:tau}
\tau = (..., s_t, a_t, \hat{G}_t,...),
\end{equation}

where $\hat{G}_t = \sum_{k=t}^{T} \gamma^{k-t} r_k$ embodies the RTG at the time $t$, highlighting the cumulative anticipated value for an agent. By its definition, RTG serves as a foresight-driven metric, quantifying the aggregated discounted rewards an agent anticipates receiving in the future. Such a metric embeds the agent's immediate decisions and future-oriented strategies, acting as a crucial linkage between the two. Hence, RTG's inclusion in our sequence modeling technique empowers the model with a prophetic viewpoint, effectively marrying the agent's short-term maneuvers with its long-term aspirations.


\subsection{Adaptive Causal Masking}
\label{masking}
In offline RL, trajectory sequences can be interpreted within the context of a sequential decision-making paradigm. Within this framework, we can consider that specific input tokens undergo masking, thereby transforming the task into an inference challenge based on particular token subsets. For a specified time \( t \), the masking configuration exposes only the tokens \( s_{0:t} \), \( a_{0:t-1} \), and \( \hat{G}_0 \). Consequently, the model is tasked with inferring the action \( a_t \), conditioned on these revealed tokens, as denoted by \( P(a_t|s_{0:t}, a_{0:t-1}, \hat{G}_0) \).

Let \( C \geq 1 \) define the context length, which corresponds to the most recent \( C \) timesteps provided to the transformer. This results in trajectory segments, \( \tau_{t-C+1:t} \), each of length \( C \). Within these segments, \( s_{t-C+1:t} \) and \( a_{t-C+1:t} \) denote sequences of the preceding \( C \) states and actions at time \( t \), respectively, while \( G_{t-C+1:t} \) signifies the RTG values over the same interval. 
Contrary to conventional methods that incorporate an RTG token at each timestep, our methodology emphasizes solely on the initial RTG token within a given context window, masking the subsequent RTGs. This minimalistic strategy, which only leverages the first RTG token, has proven effective for our inference tasks.

To enhance the agent's inference capabilities, we incorporate variability in the token length fed to the model. This variability exposes the agent to trajectory segments of diverse lengths, alternating between longer and shorter sequences. For a given segment \( \tau_{t-C+1:t} \), the masking configuration at time \( t \) is dictated by \( m \), uniformly selected from [0, \( C \)]. Consequently, the model is presented with tokens \( s_{t-m+1:t} \), \( a_{t-m+1:t-1} \), and \( \hat{G}_{t-m+1} \), while the remaining tokens are masked. And the model is tasked with predicting the action \( a_t \) for the timestep \( t \).

This adaptive causal masking approach serves a dual purpose. For smaller values of \( m \), the agent is presented with a constrained context, predominantly relying on recent states and actions. This setup fine-tunes the agent's capacity to anticipate near-future events, nurturing its short-term inference skills.
On the other hand, as \( m \) approaches \( C \), the agent is immersed in an expansive context that spans a more extended historical sequence of states and actions. This broader perspective refines the agent's capacity for long-term inference.
Through this spectrum of context lengths, the agent cultivates a versatile inference aptitude, enhancing its decision-making acumen across a range of scenarios.

\subsection{Segmented Retention Mechanism}
\label{sec_ret}
Given a sequence trajectory, the representation of states, actions, and rewards is delineated as separate tokens. This delineation triples the sequence length for a trajectory segment of length \( C \), resulting in a length of \( 3C \). Such an expansion not only amplifies computational demands but also accentuates the inherent computational bottleneck of self-attention, which inherently scales quadratically with sequence length. To address this challenge, we draw inspiration from~\cite{sun2023retentive} and introduce the multi-scale segmented retention mechanism. This segmented retention mechanism is specifically designed to replace the conventional masked multi-head attention mechanism, thereby enhancing training efficiency, especially for proficient long-sequence modeling.

Let \( d_h \) denote the size of the hidden states and \( H \in \mathbb{R}^{C \times d_h} \) represents the hidden states for the trajectory segment \( \tau_{t-C+1:t} \). We define matrices \( Q, K, V \in \mathbb{R}^{C\times d} \) as the query, key, and value matrices, respectively, where \( d \) is the embedding dimension.

The hidden state is projected into a one-dimensional function as:
\begin{equation}
    v(n) = H_n \cdot w_v,
\end{equation}
where \( w_v \) is the associated weight vector.


Central to the retention mechanism is the recurrent state \( z_n \), which captures accumulated information up to timestep \( n \). This state is pivotal in computing the output \( ret(n) \) of the mechanism, given by:
\begin{equation}
\label{z_n}
z_n = A \cdot z_{n-1} + K^{\intercal}_n \cdot v(n) , 
\end{equation}
where matrix \( A \) can be diagonalized as \( A = \Lambda( \alpha e^{i\beta} )\Lambda^{-1} \), with \( \alpha \) and \( \beta \) being vectors in \( R^d \). The output of the retention mechanism at time \( n \) is then expressed as:
\begin{equation}
\label{o_n}
ret(n) =Q_n \cdot z_n  = Q_n \cdot \sum_{m=1}^{n} A^{n-m} K^{\intercal}_m \cdot v(m).
\end{equation}

By integrating \( \Lambda \) into \( W_Q \) and \( W_K \), where \( W_Q, W_K \in R^{d_h \times d} \) are the learnable parameter matrices for the Query and Key projections, respectively, the equation becomes:
\begin{equation}
\label{rec_ot}
    ret(n) = \sum_{m=1}^{n} \left[ Q_n \cdot (\alpha e^{i\beta})^n \right] \left[ (K_m \cdot (\alpha e^{i\beta})^{-m})^{\intercal}\right] \cdot v(m).
\end{equation}

We can define:
\begin{equation}
Q = (H \cdot W_Q) \odot \omega, \quad
K = X \cdot W_K \odot  \bar{\omega}, \quad
V = X \cdot W_V,
\end{equation}
where \( \omega = e^{i\beta} \) and \( \bar{\omega} \) is the complex conjugate of \( \omega \). The output of the retention layer in a Recurrent Neural Network (RNN) manner is:
\begin{equation}
\begin{aligned} 
 & Z_n = \alpha \cdot Z_{n-1} + K^{\intercal}_n \cdot V_n \\
 & \text{Retention}_{RNN}(H_n) = Q_n \cdot Z_n
\end{aligned} 
\end{equation}

Considering \( \alpha \) as a scalar, the equation simplifies to an easily parallelizable form:
\begin{equation}
\label{ret_pa}
ret(n) = \sum_{m=1}^{n} \alpha^{n-m} \cdot \left[ Q_n \cdot \omega^n \right] \left[ (K_m \cdot \omega^m)^* \right] \cdot v(m),
\end{equation}
where \( ^* \) indicates the conjugate transpose. The output of the retention layer in a parallel manner is:
\begin{equation}
\text{Retention}_{pal}(H) = (QK^{\intercal} \odot D) \cdot V, \quad
D_{nm} =
    \begin{cases}
        \alpha^{n-m}  & n \geq m\\
        0 & n < m
    \end{cases}
\end{equation}

For handling extended sequences, we adopt a segmentation strategy. Specifically, input sequences are partitioned into segments, with parallel retention computations applied within each segment and recurrent retention computations bridging across segments. Let's assume the input sequences are divided into \( S \) segments, each of length \( M \). The retention output for the \( s \)-th segment, denoted as \( H^{(s)} \) (where \( s \in [1, S] \)), can be expressed as:
\begin{align}
  \begin{split}
 \text{Retention}\big(H^{(s)}\big) &= \text{Retention}_{pal}\big(H^{(s)}\big) +\text{Retention}_{RNN}\big(H^{(s)}\big) \\
&= \left[Q^{(s)} \big(K^{(s)}\big)^{\intercal} \odot D\right] \cdot V^{(s)} + \left[Q^{(s)}  \cdot Z^{seg}_{s-1}\right] \odot \alpha^{i+1},
   \end{split}
\end{align}
where \( Z^{\text{seg}}_{s-1} \) represents a segment-adapted version of the recurrent representation of retention, defined as:
\begin{equation}
    Z^{\text{seg}}_{s} = \alpha^M \cdot Z^{\text{seg}}_{s-1} + \big(K^{(s)}\big)^{\intercal} \cdot \big(V^{(s)} \odot \alpha^{M-i-1}\big). 
\end{equation}

\subsection{Model Architecture}

Our framework employs the RetNet architecture~\cite{sun2023retentive}, tailored for sequential modeling in offline RL for RS, incorporating adaptive causal causal masking. Comprising $L$ stacked multi-input blocks, MaskRDT processes a trajectory segment \( \tau_{t-C+1:t} \) spanning the last $C$ timesteps as input to the initial transformer block. We initiate by deriving the masked trajectory representation, masking $3C$ tokens as delineated in~\cref{masking}. 

\subsubsection{Embedding Layer}
For masked RTGs, masked states, and masked actions, we employ a linear layer to derive their respective token embeddings, which is followed by layer normalization. To convey the time-horizon information within the trajectory segment \( \tau_{t-C+1:t} \), we utilize absolute positional encoding for tokens, deviating from the conventional timestep encoding approach~\cite{NEURIPS2021_7f489f64}. This methodology curtails the propensity for overfitting often associated with direct timestep data. Additionally, we have adjusted the return-to-go token to encompass both the return value and the current timestep, ensuring the preservation of vital trajectory-level timestep insights.

\subsubsection{RetNet Decision Block}
As depicted in Figure~\ref{overrall}, our transformer is structured into $L$ consistent blocks, indexed as $l = 1,..., L$ from the bottom upwards. At each time step $t$, these blocks simultaneously generate hidden representations for the state, action, and RTG at every layer $l$.
For the masked trajectory segment \( \tau_{t-m+1:t} \), the hidden representations at layer $l$ and time step $t$ are represented as $H^l = (H^l_1, ....H^l_T)^\top \in \mathbf{R}^{T \times d_h}$. Each transformer block operates on three concurrent sequences of these representations.
The foundational block ingests the output $H^0$ from the embedding layer. For layers $l \geq2$, the input is sourced from the output of its immediate predecessor, the $(l-1)$ block:
\begin{equation}
H_t^l= \text{RetNet\_block}(H_t^{l-1}),\quad \text{for}~l\geq2.
\end{equation}
Incorporated within each block is a Multi-Head Retention, which acts on the input tokens and is followed by a Position-wise Feed-Forward layer.

\noindent\textbf{Multi-Head Retention Mechanism.}
Consider \( h = d_h/d \), which denotes the number of attention heads. As detailed in Section~\cref{sec_ret}, each attention head produces a representation described by:
\begin{equation}
    \text{head}^j = \text{Retention}(H_t^l, \alpha_j), \quad
    \alpha = 1 - 2^{-5 - \text{linspace}(0, h-1, h)} \in \mathbb{R}^h.
\end{equation}
Each attention head is assigned a unique \( \alpha \) value. The multi-head retention mechanism operates by applying \( h \) attention functions in parallel to create a unified output representation. This output is a concatenated projection of the representations from all the heads:
\begin{equation}
    \text{MultiHead} = \text{Concat}(\text{head}^1, ..., \text{head}^h).
\end{equation}
To ensure consistent scaling and normalization across different attention heads, we employ Group Normalization (GN)~\cite{wu2018group} on the outputs of each head.
To enhance the non-linearity within the retention layers, we incorporate the Swish activation function~\cite{ramachandran2017searching}. 
The final multi-head retention output, denoted as \( MHR(X) \), is formulated as:
\begin{equation}
    MSR(H^l_t) = \left[ (H^l_t W_P \cdot \text{sigmoid}(H^l_t W_P)) \odot \text{GN(MultiHead)}\right]W_O,
\end{equation}
where \( W_P \), \( W_O \in R^{d_h \times d_h} \) and represent learnable parameters.

\noindent\textbf{Position-wise Feed-Forward Networks.}
Within each block, to enhance the representations derived from the retention layer and adapt them to the specific task at hand, we incorporate a Position-wise Feed-Forward Network (FFN). This FFN is delineated by two linear layers, interspersed with an activation function. The resulting output is formulated as:
\begin{equation}
    \text{FFN}(MSR(H^l_t)) = \text{GELU}\big((MSR(H^l_t) W_1 + b_1\big)W_2+b_2\big),
\end{equation}
where the activation function employed is the Gaussian Error Linear Unit (GELU)~\cite{hendrycks2016bridging}.

\subsubsection{Causal Layer.}
The culmination of the $L$ RetNet blocks yields an output represented as $H^L_t = (s^L_{t-m+1}, a^L_{t-m+1}, G^L_{t-m+1}, ..., s^L_t, -, -)$, where the symbol '-' denotes masked tokens. This output is segmented into three distinct sets: RTG, states, and actions. This can be articulated as:
\begin{equation}
    H^L_t = (s_{t-m+1:t}, a_{t-m+1:t-1}, G_{t-m+1}).
\end{equation} 

Given that the action inference is contingent upon all unmasked tokens, the final representation is synthesized from these three hidden states. To achieve action prediction, we apply a linear layer followed by the GELU activation:
\begin{gather}
        \tilde{\Psi}^a_t = s_{t-m+1:t} + a_{t-m+1:t-1} + G_{t-m+1}\\
    \Psi^a_t = \text{GELU}(\tilde{\Psi}^a_t W+b)
\end{gather}

In a parallel vein, the prediction for the subsequent state is derived from $s_{t-K+1:t}$ and $\Psi^a_t$, as state transitions inherently depend on the preceding state and the current action:
\begin{gather}
        \tilde{\Psi}^s_t = s_{t-K+1:t} + \Psi^a_t\\
    \Psi^s_t = \text{GELU}(\tilde{\Psi}^s_t W+b)
\end{gather}

To mitigate overfitting, we introduce Dropout~\cite{srivastava2014dropout} post the linear layer's output. Both the action and state predictions are subsequently processed through fully-connected networks, termed the reward estimation network $N_e$, to deduce the prospective reward $r_t$. This potential reward, alongside the action prediction, is then channeled into the action generation network $N_g$ to produce the final anticipated action $a_t$.

\subsection{Training Procedure}

Our training procedure is grounded on a dataset of recommendation trajectories. Initially, the Deep Deterministic Policy Gradient (DDPG) method~\cite{lillicrap2015continuous} is employed to train an expert reinforcement learning (RL) agent. This expert agent then interacts with the environment to produce a collection of expert trajectories, which form our primary dataset.

From this dataset, we sample mini-batches of sequences, each characterized by a context length of \(C\). These sequences undergo a transformation to be suitable as network inputs. After this transformation, a masking configuration is applied to the input, which is then processed through the entirety of the network blocks.

Within our model, \(\theta_e\) represents the trainable parameters for the reward estimation network \(N_e\), \(\theta_g\) denotes those for the action generation network \(N_g\), \(\theta_s\) encapsulates the parameters for generating the state representation \(\Psi^s_t\), and \(\theta_a\) is reserved for generating the action representation \(\Psi^a_t\).

The reward estimation network \(N_e\), in conjunction with state and action predictions, is optimized by minimizing the factual reward loss:
\begin{equation}
\mathcal{L}_{N_e}(\theta_e, \theta_s, \theta_a) = \mathbb{E}_{(s,a,G)\sim \tau}\left[\frac{1}{K}\Sigma_{k=1}^K\left(r_k - N_e\big(\Psi^s_t(\theta_s), \Psi^a_t(\theta_a);\theta_e\big)\right)^2\right].
\end{equation}

For the action generation network \(N_g\), the objective is to produce the final action by minimizing its cross-entropy loss:
\begin{equation}
\mathcal{L}_{N_g}(\theta_g, \theta_e, \theta_a) = \frac{1}{K}\mathbb{E}_{(s,a,G)\sim \tau}\left[-\Sigma_{k=1}^K \log N_g\big(\Psi^a_t(\theta_a);\theta_g \big)\right].
\end{equation}

The overarching loss function amalgamates the losses mentioned above:
\begin{equation}
\mathcal{L} = \mathcal{L}_{N_e}(\theta_e, \theta_s, \theta_a) + \beta \mathcal{L}_{N_g}(\theta_g, \theta_e, \theta_a),
\end{equation}
where \(\beta\) serves as a hyper-parameter, modulating the contributions of the individual losses.

\section{Experiments}

In this section, we present experimental results addressing four primary research questions:

\begin{itemize}
    \item \textbf{RQ1}: How does the performance of MaskRDT stack up against traditional deep RL algorithms in both online recommendation and offline dataset environments?
    \item \textbf{RQ2}: What is the impact of adaptive causal masking on performance across different context lengths?
    \item \textbf{RQ3}: How does MaskRDT influence training cost?
    \item \textbf{RQ4}: How is the performance of MaskRDT affected by varying dataset sizes?
\end{itemize}

We focus our investigations on RQ2, RQ3 and RQ4 within online simulation settings, as they more accurately mirror real-world scenarios. Offline datasets, being static, don't capture the dynamic nature of users' interests.

\subsection{Experimental Setup: Datasets and Simulation Environments}
In this subsection, we detail the datasets and environments employed to evaluate the performance of our MaskRDT algorithm in comparison with other leading methods. Our model is implemented using PyTorch and all experiments are executed on a server equipped with two Intel Xeon CPU E5-2697 v2 CPUs, six NVIDIA TITAN X Pascal GPUs, two NVIDIA TITAN RTX GPUs, and 768 GB of RAM.


\subsubsection{Dataset} We utilize six diverse, publicly available datasets from various recommendation domains for our offline evaluations, each exhibiting distinct levels of sparsity:

\begin{itemize}
    \item \textbf{GoodReads}: A dataset from the book review platform \textit{GoodReads}\footnote{https://www.goodreads.com/}~\cite{wan2018item}, which includes varied user interactions with books such as ratings and reviews. 
    \item \textbf{LibraryThing}: Originating from \textit{LibraryThing}\footnote{https://www.librarything.com/}, a digital service aiding users in book cataloging. This dataset assists in cataloging books and captures social networking features, making it suitable for studying social-based recommendation models.
    \item \textbf{Netflix}: A renowned dataset from the Netflix Prize Challenge\footnote{https://www.kaggle.com/datasets/netflix-inc/netflix-prize-data}, is a collection of movie ratings used for recommendation system research.
    \item \textbf{Amazon CD}\footnote{https://nijianmo.github.io/amazon/index.html}: This dataset is a subset of product reviews from Amazon.com, specifically focusing on the "CD" category~\cite{ni2019justifying}.
    \item \textbf{MovieLens}: 
    A standard dataset for benchmarking recommender systems. Our study employs two versions of MovieLens datasets: MovieLens-1M\footnote{https://grouplens.org/datasets/movielens/1m/} and MovieLens-20M\footnote{https://grouplens.org/datasets/movielens/20m/}, which differ in scale.
\end{itemize}

To facilitate reinforcement learning interactions, we transform these offline datasets into simulated environments, drawing inspiration from prior works~\cite{chen2020knowledge,zhao2018recommendations}. We employ an LSTM-based state encoder to capture temporal dynamics within the data.

\subsubsection{Online Simulator} 
We conduct experiments on the VirtualTB~\cite{shi2019virtual} online simulation platform. VirtualTB replicates the dynamics of an online retail environment, having been trained on real data from the Taobao platform, one of China's premier online retail platforms. The VirtualTaobao interacts with customers by first sampling a feature vector that includes the customer's profile and search query. It then retrieves a set of items related to the query and uses a model to assign a weight vector to these items' attributes. The system calculates the dot product of this weight vector with each item's attributes, selecting the top 10 items based on these values. These items are presented to the customer, who can either click on them, navigate to the next page (prompting the system to update customer features and reiterate the process), or exit the platform.

\subsubsection{Evaluation Metric}
In the simulated online environment, we employ the Click-Through Rate (CTR) as the primary metric for assessing the performance of Reinforcement Learning Recommender Systems (RLRS). The CTR is calculated using the formula:
\begin{equation}
\text{CTR} = \frac{\text{episode\_return}}{\text{episode\_length} \times \text{maximum\_reward}},
\end{equation}
where $episode\_return$ represents the total reward accumulated in an episode, $episode\_length$ denotes the number of steps in the episode, and $maximum\_reward$ is the highest reward achievable in a single step.

For offline datasets, our evaluation encompasses a range of metrics, including recall, precision, and normalized discounted cumulative gain (nDCG), which are among the most popular metrics utilized by RLRSs, due to the absence of metrics specifically developed for RLRSs~\cite{afsar2022reinforcement}.



\subsection{Baselines}
Most of the existing works are evaluating their methods on offline datasets, and very few works provide a public online simulator evaluation. 
As there are two types of experiments, we provide two sets of 
baselines to be used for different experimental settings. Firstly, we will introduce the baselines for the online simulator, which are probably the most popular benchmarks in reinforcement learning:
\begin{itemize}
    \item \textbf{Deep Deterministic Policy Gradient (DDPG)}~\cite{lillicrap2015continuous} is an off-policy method for environments with continuous action spaces. DDPG employs a target policy network to compute an action that approximates maximization to deal with continuous action spaces.
    \item \textbf{Soft Actor Critic (SAC)}~\cite{haarnoja2018soft} is an off-policy maximum entropy Deep Reinforcement Learning approach that optimizes a stochastic policy. It employs the clipped double-Q method and entropy regularisation that trains the policy to maximize a trade-off between expected return and entropy.
    \item \textbf{Twin Delayed DDPG (TD3)}~\cite{fujimoto2018addressing} is an algorithm that improves baseline DDPG performance by incorporating three key tricks: learning two Q-functions instead of one, updating the policy less frequently, and adding noise to the target action.
    \item \textbf{Decision Transformer (DT)}~\cite{NEURIPS2021_7f489f64} is an offline reinforcement learning algorithm that incorporates the transformer as the major network component to infer actions.
\end{itemize}
Moreover, the following recommendation algorithms are used for offline evaluations which come from two different categories: transformer-based methods and reinforcement learning-based methods.
\begin{itemize}
    \item \textbf{SASRec}~\cite{kang2018self} is a well-known baseline that uses the self-attention mechanism to make sequential recommendations.
    \item \textbf{BERT4Rec}~\cite{sun2019bert4rec} is a recent transformer based method for recommendation. It adopts BERT to build a recommender system.
    \item \textbf{S3Rec}~\cite{zhou2020s3} is BERT4Rec follow-up work that uses transformer architecture and self-supervised learning to maximize mutual information.
    \item \textbf{KGRL}~\cite{chen2020knowledge} is a reinforcement learning-based method that utilizes the capability of Graph Convolutional Network (GCN) to process the knowledge graph information.
    \item \textbf{TPGR}~\cite{chen2019large} is a model that uses reinforcement learning and binary tree for large-scale interactive recommendations.
    \item \textbf{PGPR}~\cite{xian2019reinforcement} is a knowledge-aware model that employs reinforcement learning for explainable recommendations.
    \item \textbf{CDT4Rec}~\cite{Wang2023} is a casual decision transformer model for offline RLRS.
\end{itemize}
We note that SASRec, BERT4Rec, and S3Rec are not suitable for the reinforcement learning evaluation procedure.
In order to evaluate the performance of those models, we feed the trajectory representation $\tau$ as an embedding into those models for training purposes and use the remaining trajectories for testing purposes.
\begin{figure}[!h]
    \centering
    \includegraphics[width=0.9\linewidth]{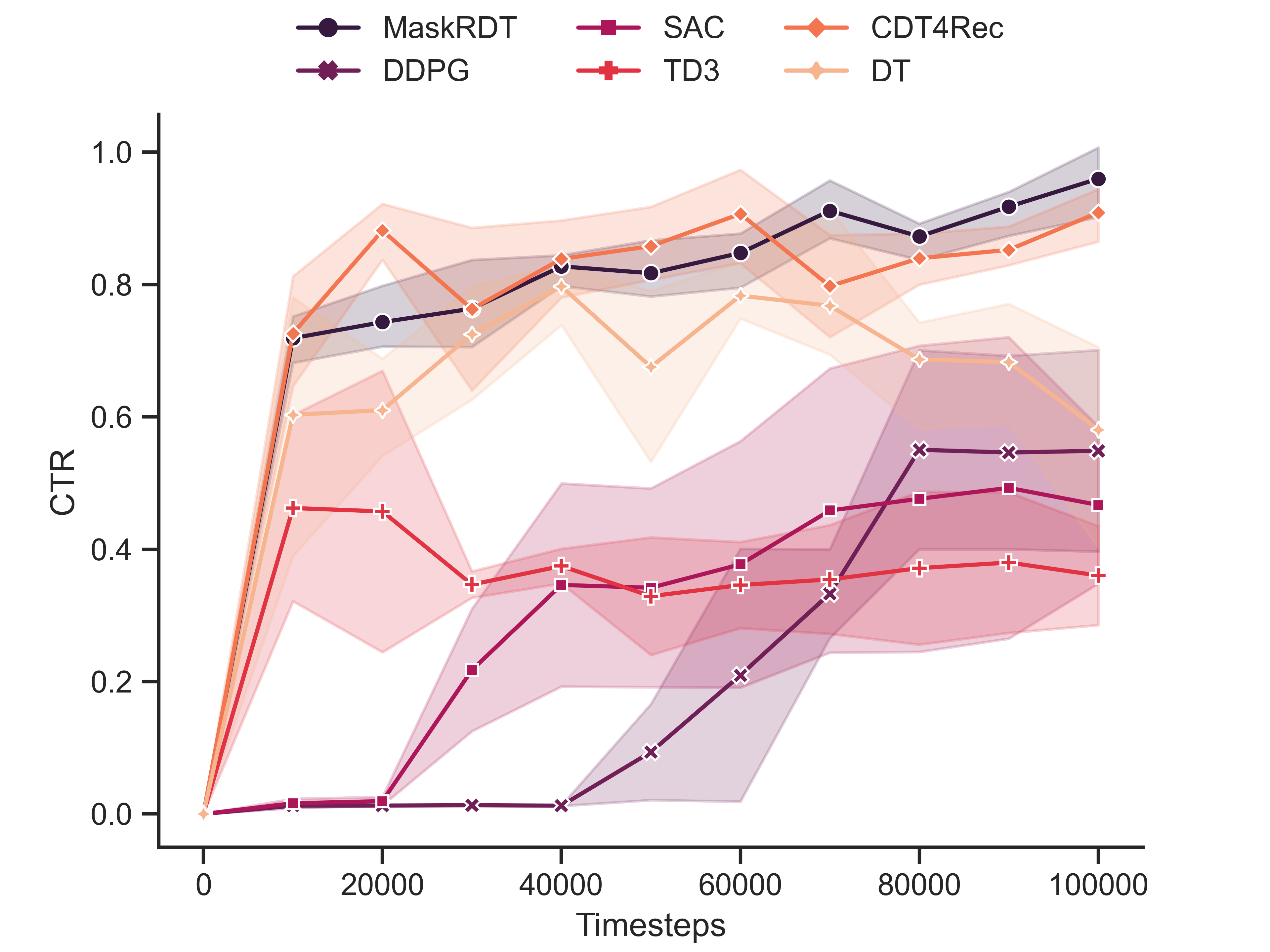}
    \caption{Overall comparison result with variance between the baselines and CDT4Rec in the VirtualTaobao simulation environment.}
    \label{fig:over_comp}
\end{figure}



\begin{table*}[!ht]
\caption{The overall results of our model comparison with several state-of-the-art models on different datasets. The highest results are in bold and the second highest are marked with underline. * indicates the statistically significant improvements (i.e., two-sided t-test with $p$ < 0.05) over the best baseline.}\smallskip
\begin{minipage}[ht]{1.0\linewidth}
\resizebox{\columnwidth}{!}{%
\begin{tabular}{cccc|ccc}
\hline
\multirow{2}{*}{Measure (\%)} & \multicolumn{3}{c|}{GoodReads}                                                       & \multicolumn{3}{c}{Librarything}                                                     \\ \cline{2-7} 
                              & Recall                      & Precision                  & nDCG                      & Recall                      & Precision                  & nDCG                      \\ \hline
SASRec                        & 6.921 $\pm$ 0.312           & 5.242 $\pm$ 0.211          & 6.124 $\pm$ 0.210         & 8.312 $\pm$ 0.201           & 6.526 $\pm$ 0.129          & 7.391 $\pm$ 0.201         \\
BERT4Rec                      & 8.483 $\pm$ 0.234           & 7.817 $\pm$ 0.281          & 8.012 $\pm$ 0.199         & 11.982 $\pm$ 0.123          & 9.928 $\pm$ 0.201          & 10.021 $\pm$ 0.210        \\
S3Rec                         & 10.263 $\pm$ 0.212          & 9.726 $\pm$ 0.188          & 10.002 $\pm$ 0.210        & 13.425 $\pm$ 0.182          & 11.725 $\pm$ 0.182         & 11.237 $\pm$ 0.127        \\
KGRL                          & 7.459 $\pm$ 0.401           & 6.444 $\pm$ 0.321          & 7.331 $\pm$ 0.301         & 12.128 $\pm$ 0.241          & 12.451 $\pm$ 0.242         & 13.925 $\pm$ 0.252        \\
TPGR                          & 11.219 $\pm$ 0.323          & 10.322 $\pm$ 0.442         & 9.825 $\pm$ 0.642         & 14.713 $\pm$ 0.644          & 12.410 $\pm$ 0.612         & 13.225 $\pm$ 0.722        \\
PGPR                          & 11.421 $\pm$ 0.223          & 10.042 $\pm$ 0.212         & 9.234 $\pm$ 0.242         & 11.531 $\pm$ 0.241          & 10.333 $\pm$ 0.341         & 12.641 $\pm$ 0.442        \\
CDT4Rec                       & \underline{13.274 $\pm$ 0.287}    & \underline{11.276 $\pm$ 0.175}   & \underline{10.768 $\pm$ 0.372}  & \underline{ 15.229 $\pm$ 0.128}    & \underline{ 14.020 $\pm$ 0.201}   & \underline{ 14.768 $\pm$ 0.176}  \\
MaskRDT(Ours)                       & \textbf{13.512 $\pm$ 0.224}* & \textbf{11.404 $\pm$0.198}* & \textbf{10.934$\pm$ 0.252}* & \textbf{15.403 $\pm$ 0.133}* & \textbf{14.152 $\pm$0.198}* & \textbf{14.855$\pm$0.159}* \\ \hline
\end{tabular}
}
\end{minipage}

\begin{minipage}[ht]{1.0\linewidth}
\resizebox{\columnwidth}{!}{%
\begin{tabular}{cccc|ccc}
\hline
\multirow{2}{*}{Measure (\%)} & \multicolumn{3}{c|}{Amazon CD}                                                            & \multicolumn{3}{c}{Netflix}                                                                \\ \cline{2-7} 
                              & Recall                        & Precision                   & nDCG                        & Recall                        & Precision                    & nDCG                        \\ \hline
SASRec                        & 5.210 $\pm$ 0.202             & 2.352 $\pm$ 0.124           & 4.601 $\pm$ 0.282           & 11.321 $\pm$ 0.231            & 10.322 $\pm$ 0.294           & 14.225 $\pm$ 0.421          \\
BERT4Rec                      & 9.123 $\pm$ 0.200             & 6.182 $\pm$ 0.211           & 7.123 $\pm$ 0.198           & 13.847 $\pm$ 0.128            & 12.098 $\pm$ 0.256           & 13.274 $\pm$ 0.210          \\
S3Rec                         & 10.212 $\pm$ 0.192            & 7.928 $\pm$ 0.222           & 8.028 $\pm$ 0.129           & 14.090 $\pm$ 0.227            & 12.349 $\pm$ 0.256           & 13.002 $\pm$ 0.281          \\
KGRL                          & 8.208 $\pm$ 0.241             & 4.782 $\pm$ 0.341           & 6.876 $\pm$ 0.511           & 13.909 $\pm$ 0.343            & 11.874 $\pm$ 0.232           & 13.082 $\pm$ 0.348          \\
TPGR                          & 7.294 $\pm$ 0.312             & 2.872 $\pm$ 0.531           & 6.128 $\pm$ 0.541           & 12.512 $\pm$ 0.556            & 11.512 $\pm$ 0.595           & 10.425 $\pm$ 0.602          \\
PGPR                          & 6.619 $\pm$ 0.123             & 1.892 $\pm$ 0.143           & 5.970 $\pm$ 0.131           & 10.982 $\pm$ 0.181            & 10.123 $\pm$ 0.227           & 10.104 $\pm$ 0.243          \\
CDT4Rec                       & \underline{10.424 $\pm$ 0.122}      & \underline{8.212 $\pm$ 0.201}           & \underline{8.111 $\pm$ 0.182}           & \underline{15.229 $\pm$ 0.128}            & \underline{14.020 $\pm$ 0.201}           & \underline{14.768 $\pm$ 0.176}          \\
MaskRDT(Ours)                 & \textbf{10.522 $\pm$ 0.103 *} & \textbf{8.398 $\pm$0.167 *} & \textbf{8.323$\pm$ 0.159 *} & \textbf{15.372 $\pm$ 0.122 *} & \textbf{14.205 $\pm$0.133 *} & \textbf{14.902$\pm$0.133 *} \\ \hline
\end{tabular}
}
\end{minipage}

\begin{minipage}[ht]{1.0\linewidth}
\resizebox{\columnwidth}{!}{%
\begin{tabular}{cccc|ccc}
\hline
\multirow{2}{*}{Measure (\%)} & \multicolumn{3}{c|}{MovieLens-1M}                                                       & \multicolumn{3}{c}{MovieLens-20M}                                                       \\ \cline{2-7} 
                              & Recall                       & Precision                    & nDCG                      & Recall                      & Precision                   & nDCG                        \\ \hline
SASRec                        & 5.831 $\pm$ 0.272            & 2.352 $\pm$ 0.124            & 4.601 $\pm$ 0.282         & 14.512 $\pm$ 0.510          & 12.412 $\pm$ 0.333          & 12.401 $\pm$ 0.422          \\
BERT4Rec                      & 8.222 $\pm$ 0.192            & 6.182 $\pm$ 0.211            & 7.123 $\pm$ 0.198         & 17.212 $\pm$ 0.233          & 14.234 $\pm$ 0.192          & 13.292 $\pm$ 0.212          \\
S3Rec                         & 8.992 $\pm$ 0.265            & 7.928 $\pm$ 0.222            & 8.028 $\pm$ 0.129         & 17.423 $\pm$ 0.128          & 15.002 $\pm$ 0.221          & 13.429 $\pm$ 0.520          \\
KGRL                          & 8.004 $\pm$ 0.223            & 4.782 $\pm$ 0.341            & 6.876 $\pm$ 0.511         & 16.021 $\pm$ 0.498          & 14.989 $\pm$ 0.432          & 13.007 $\pm$ 0.543          \\
TPGR                          & 7.246 $\pm$ 0.321            & 2.872 $\pm$ 0.531            & 6.128 $\pm$ 0.541         & 16.431 $\pm$ 0.369          & 13.421 $\pm$ 0.257          & 13.512 $\pm$ 0.484          \\
PGPR                          & 6.998 $\pm$ 0.112            & 1.892 $\pm$ 0.143            & 5.970 $\pm$ 0.131         & 14.234 $\pm$ 0.207          & 9.531 $\pm$ 0.219           & 11.561 $\pm$ 0.228          \\
CDT4Rec                       & \underline{9.234 $\pm$ 0.123}            & \underline{8.212 $\pm$ 0.201}            & \underline{8.111 $\pm$ 0.182}        & \underline{19.273 $\pm$ 0.212}          & \underline{17.371 $\pm$ 0.276}          & \underline{17.311 $\pm$ 0.216}          \\
MaskRDT(Ours)                       & \textbf{9.326 $\pm$ 0.131 *} & \textbf{8.323 $\pm$ 0.173 *} & \textbf{8.254$\pm$0.195*} & \textbf{19.396$\pm$0.179 *} & \textbf{17.501$\pm$0.204 *} & \textbf{17.445$\pm$0.241 *} \\ \hline
\end{tabular}
}
\end{minipage}
\label{tab:result}
\end{table*}

\begin{figure*}[h]
  \begin{subfigure}{0.28\textwidth}
    \centering
    \includegraphics[width=\linewidth]{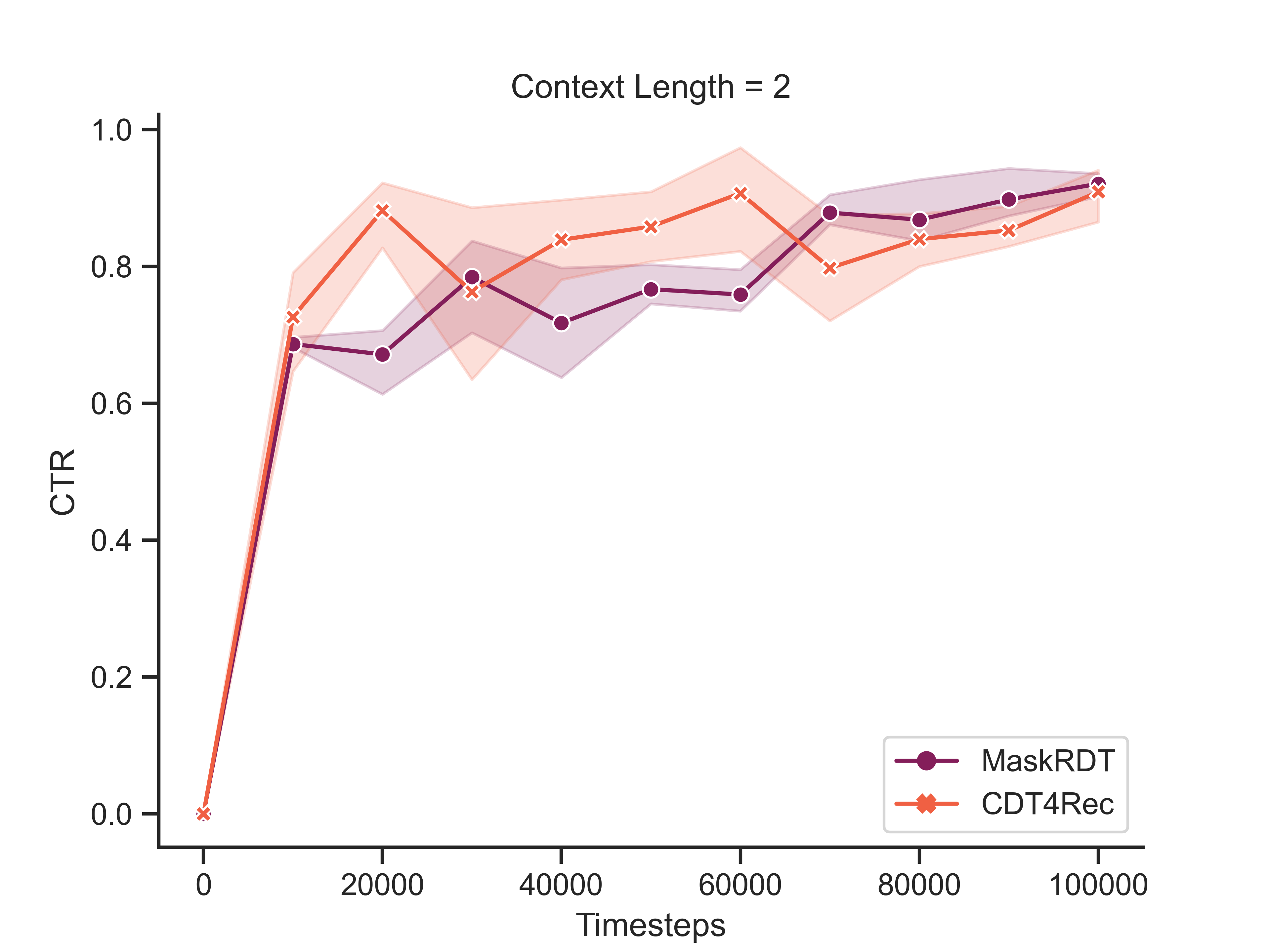}
    \label{fig:1}
  \end{subfigure}%
  \begin{subfigure}{0.28\textwidth}
    \centering
    \includegraphics[width=\linewidth]{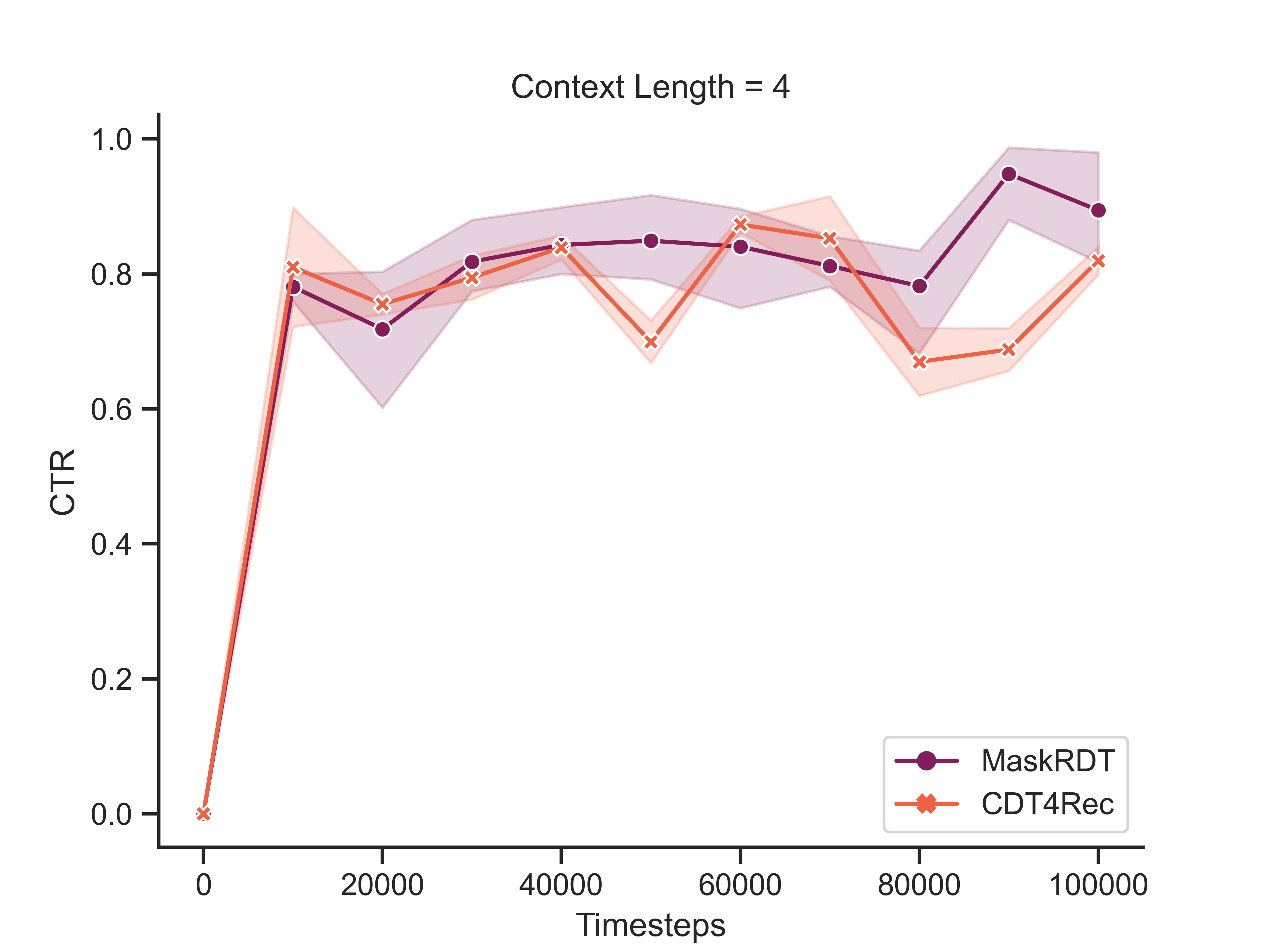}
    \label{fig:2}
  \end{subfigure}
  \begin{subfigure}{0.28\textwidth}\quad
    \centering
    \includegraphics[width=\linewidth]{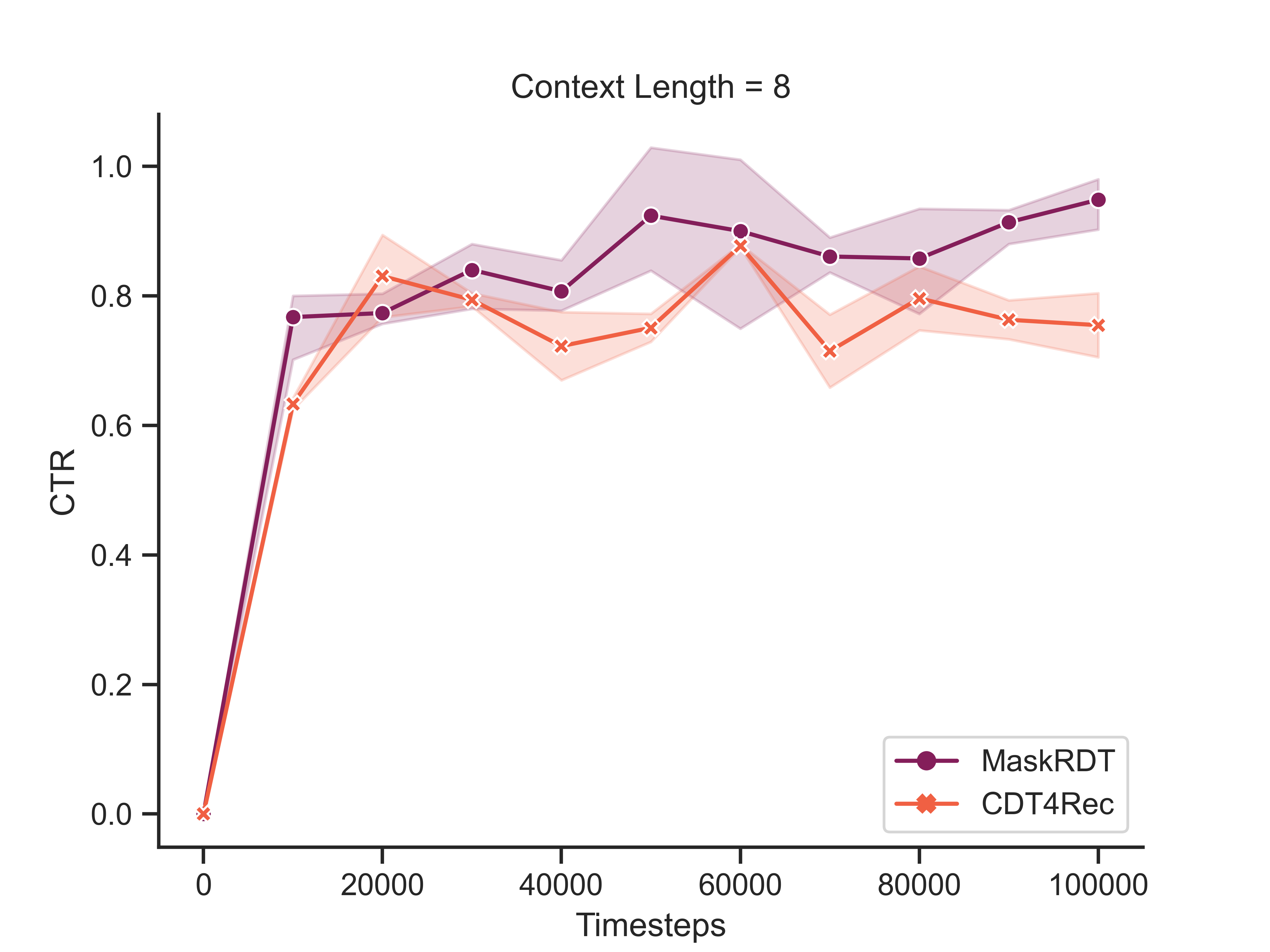}
    \label{fig:3}
  \end{subfigure}
  \medskip

  \begin{subfigure}{0.28\textwidth}
    \centering
    \includegraphics[width=\linewidth]{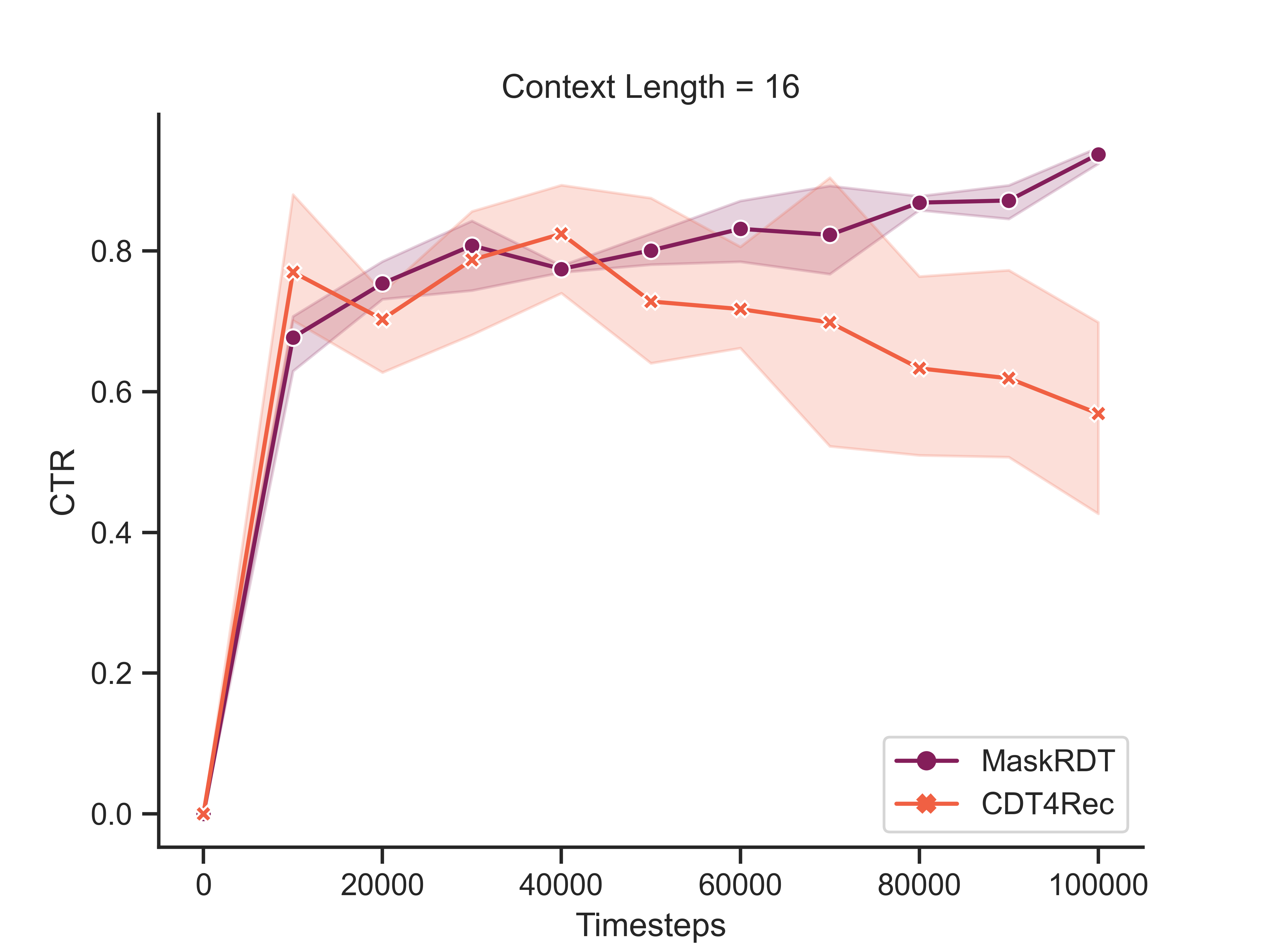}
    \label{fig:4}
  \end{subfigure}
  \begin{subfigure}{0.28\textwidth}
    \centering
    \includegraphics[width=\linewidth]{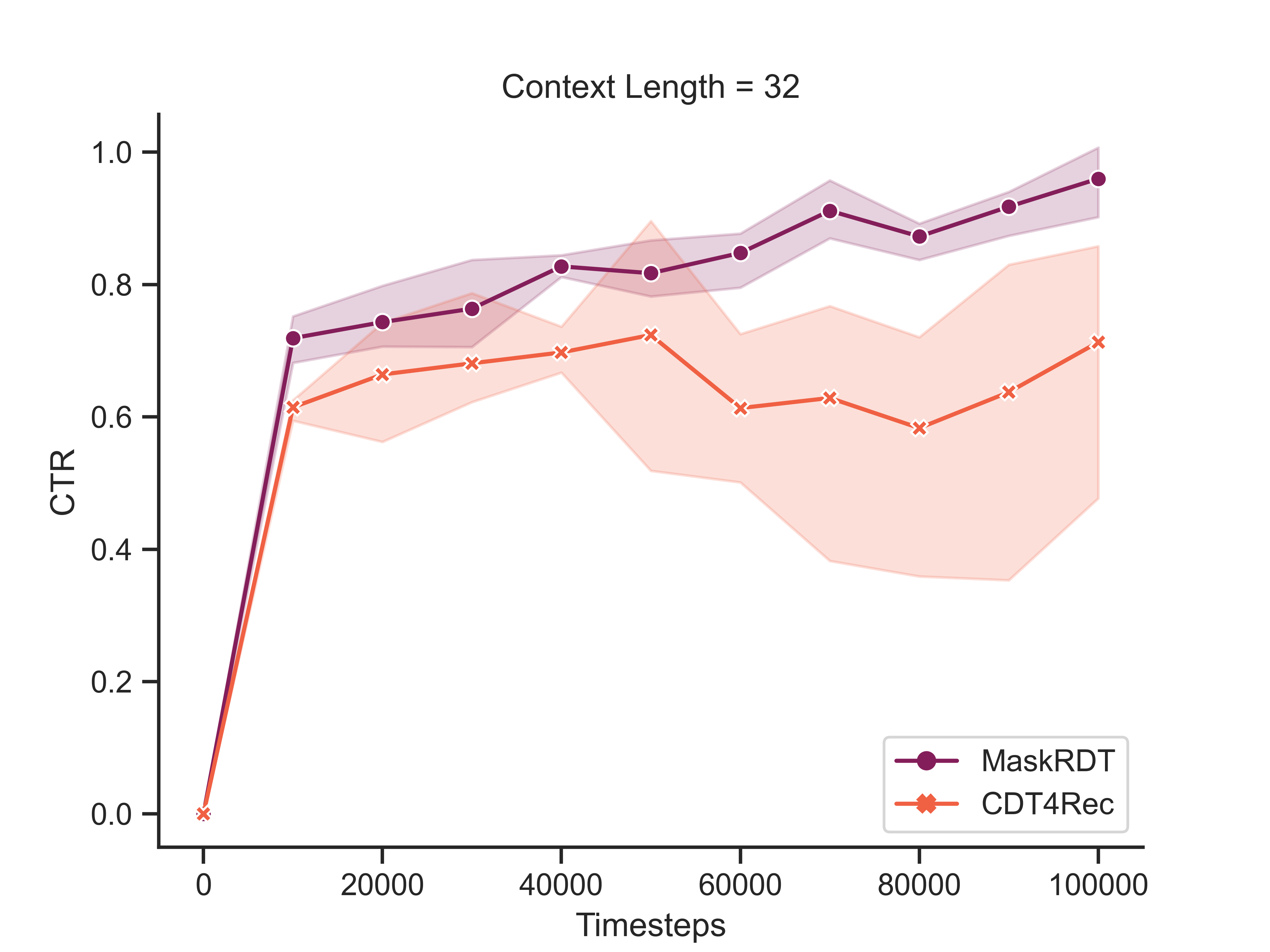}
    \label{fig:5}
  \end{subfigure}
  \begin{subfigure}{0.28\textwidth}
    \centering
    \includegraphics[width=\linewidth]{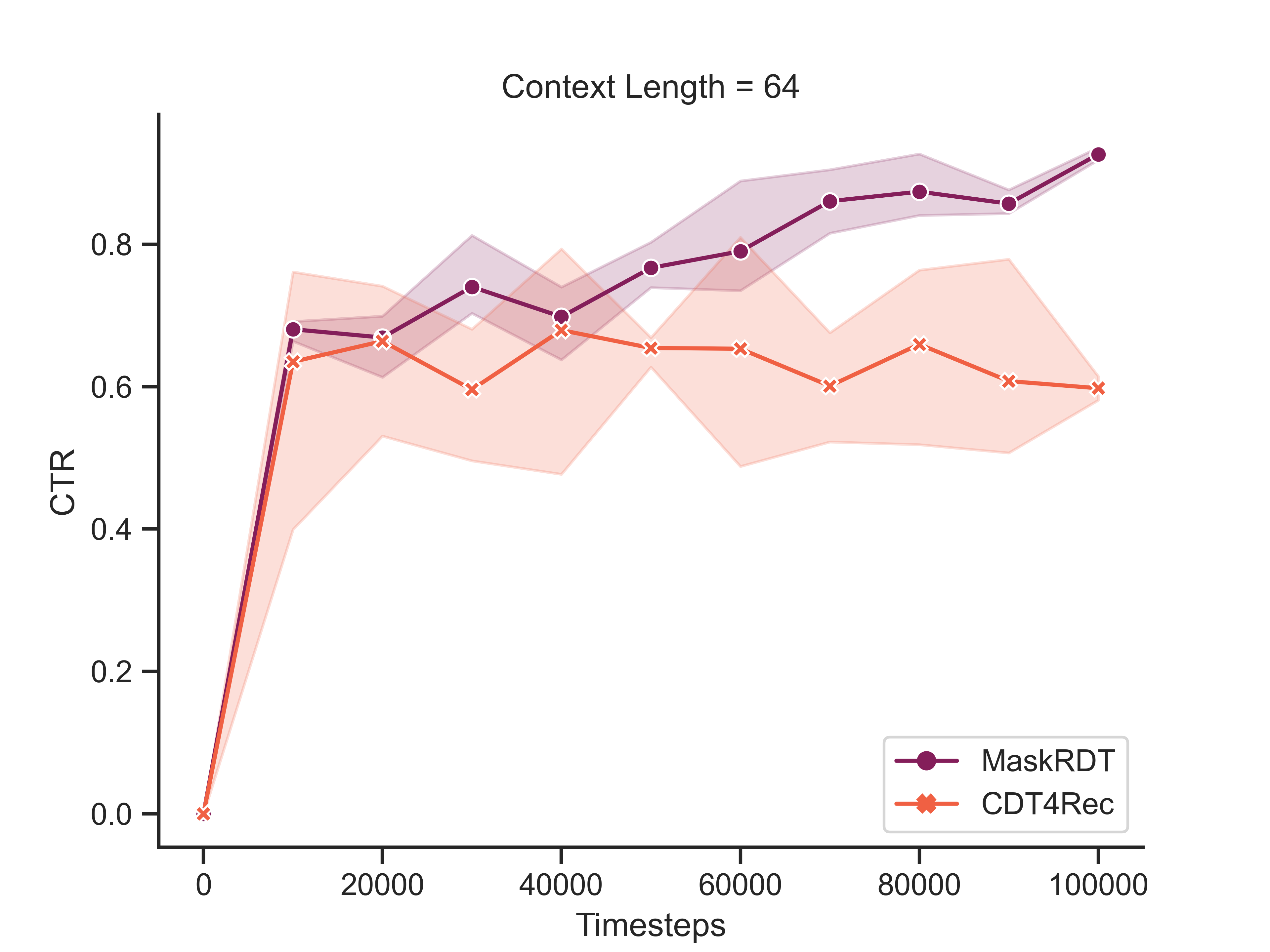}
    \label{fig:6}
  \end{subfigure}
  \caption{Performance Comparison Between MaskRDT and CDT4Rec for Different Context Lengths}
  \label{fig:images}
\end{figure*}


\subsection{Overall Results in the Online Simulator (RQ1)}

This section initiates with a description of our methodology for conducting experiments within an online simulator, employing offline RL techniques. Our approach begins with the training of an expert agent through DDPG, which is subsequently deployed within the simulator to accumulate expert trajectories. It is imperative to note that the expert's interaction with the environment is intentionally restricted, with the goal of collectiong a specific number of random trajectories. These collected trajectories constitute the initial dataset, which is essential for the pre-training phase of offline RL methods. Subsequently, the offline RL algorithm is refined through interactions within the simulated environment.

The effectiveness of our MaskRDT method, in comparison to baseline approaches, is delineated in Figure~\ref{fig:over_comp}, which demonstrates the CTR performance across iterative timesteps in the VirtualTaobao simulation. The data reveals a marked improvement in performance, with MaskRDT outperforming nearly all competing algorithms around the 20,000-timestep threshold.As the simulation progresses to 100,000 timesteps, MaskRDT's CTR stabilizes at a plateau approximately 0.9, reflecting its potent recommendation capabilities. The variance in our model's CTR, illustrated by the shaded region around the MaskRDT trajectory, is narrower than that of the competing methods, underscoring our model's consistent learning and dependability.

When compared with baselines, it becomes apparent that CDT4Rec, while following a rising trend similar to MaskRDT, experiences a higher variance and falls short of MaskRDT's peak CTR. Other algorithms like SAC and TD3 achieve moderate success but do not reach the high consistency level displayed by MaskRDT. In contrast, DDPG and DT show significantly lower effectiveness in this simulated setting.

In summary, within the VirtualTaobao online simulation, MaskRDT exhibits a pronounced superiority in maximizing CTR, suggesting its potential as a more efficient system for providing clickable recommendations to users in comparison to the evaluated baselines.

\subsection{Overall Results on Offline Dataset(RQ1)}
The performance of our MaskRDT method against baseline models across several offline datasets is summarized in~\Cref{tab:result}. MaskRDT consistently superior its counterparts in measures of Recall, Precision, and nDCG across datasets such as GoodReads, Librarything, Amazon CD, Netflix, MovieLens-1M, and MovieLens-20M.

Particularly, MaskRDT achieves notable high scores in the Good-Reads and Librarything datasets with the highest nDCG values of 10.934\% and 14.855\% respectively. It continues this leading trend in Amazon CD and Netflix datasets, showing superior Recall and Precision. In the extensive MovieLens-1M and 20M datasets, MaskRDT maintains its dominance with the highest Recall and nDCG scores, emphasizing its robust recommendation capabilities.

The results emphasize how effective and stable MaskRDT is in offline settings, showcasing its ability to learn from different context lengths as needed. This adaptability is crucial in allowing our model to find a delicate balance between understanding recent user behavior trends and respecting long-standing preferences, ultimately boosting the relevance and precision of its recommendations.
\begin{figure*}[h]
     \centering
     \begin{subfigure}[b]{0.28\linewidth}
         \centering
         \includegraphics[width=\linewidth]{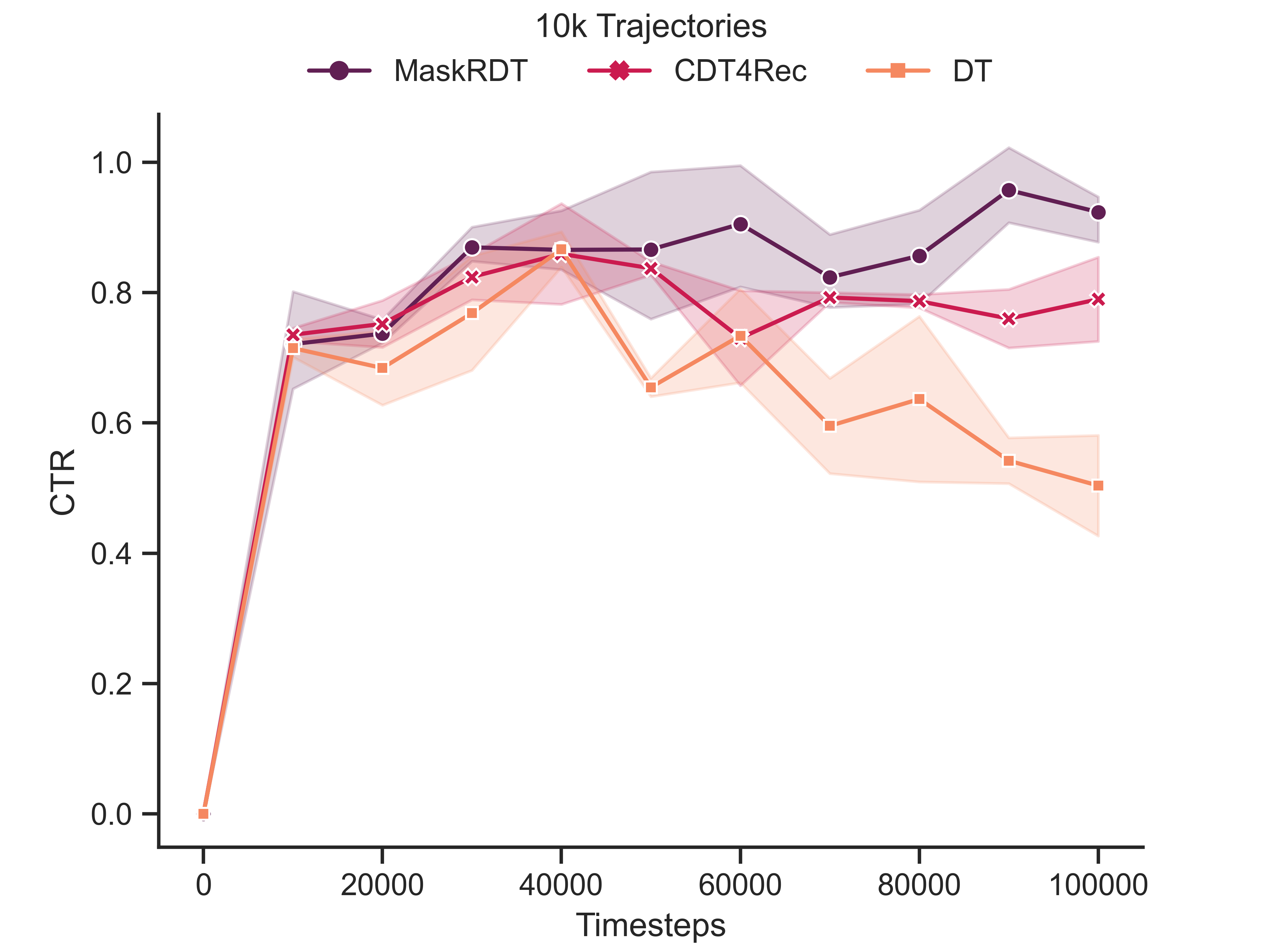}
     \end{subfigure}
     \begin{subfigure}[b]{0.28\linewidth}
         \centering
         \includegraphics[width=\linewidth]{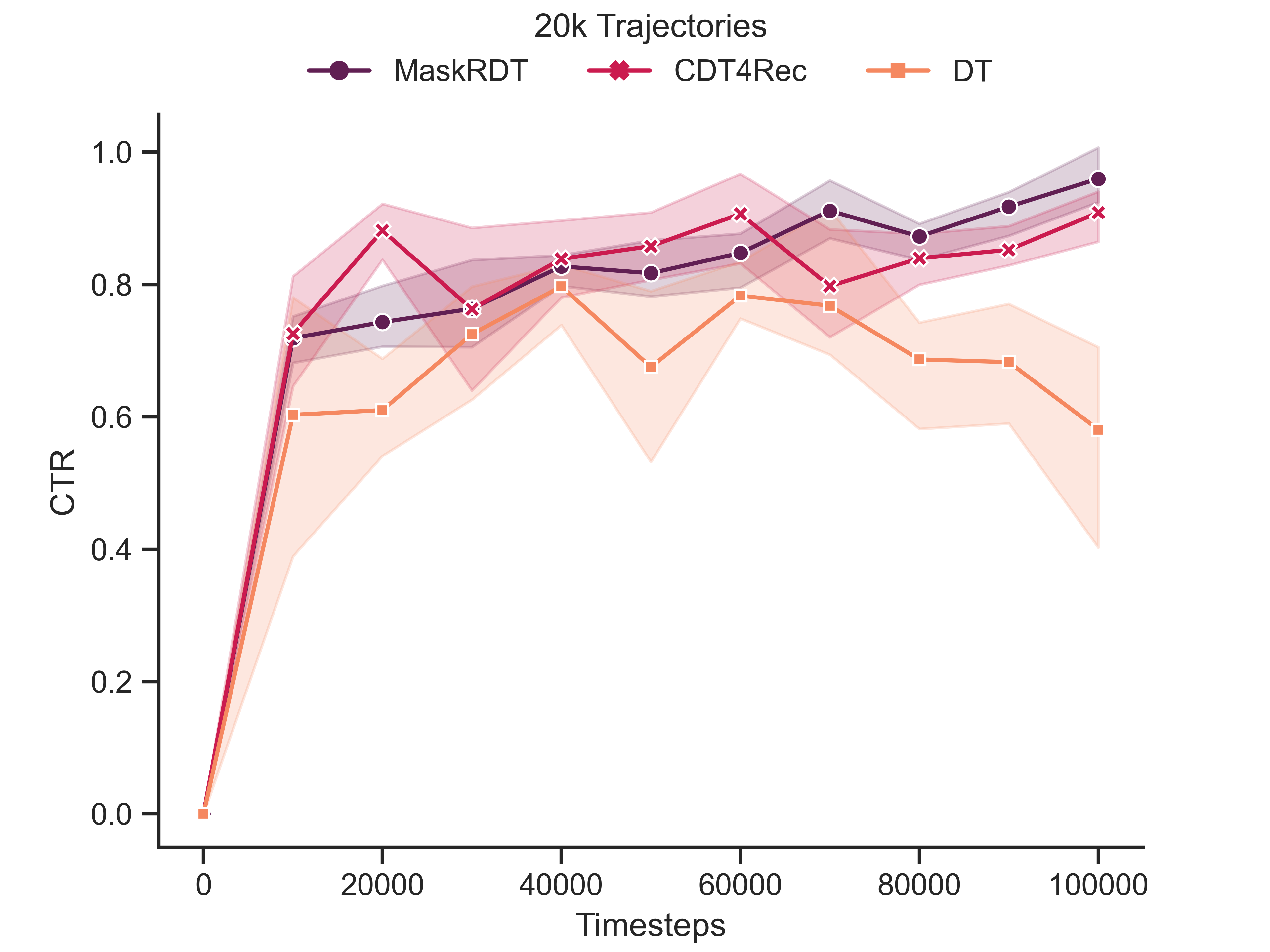}
     \end{subfigure}
     \begin{subfigure}[b]{0.28\linewidth}
         \centering
         \includegraphics[width=\linewidth]{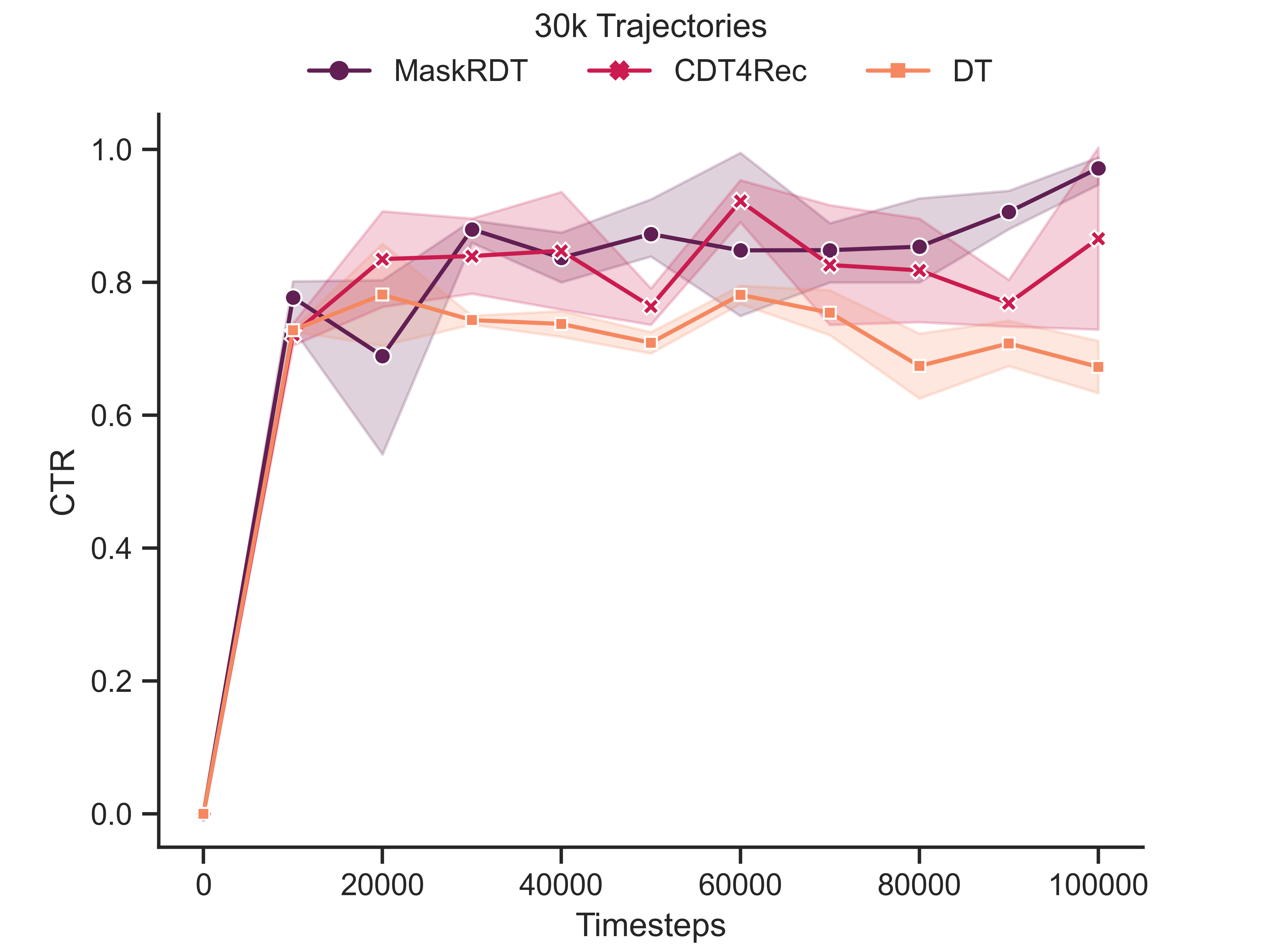}
     \end{subfigure}
        \caption{Performance Comparison Between MaskRDT, CDT4Rec and DT for Different Numbers of Trajectories}
\label{fig:num_of_traje}
\end{figure*}

\subsection{Ablation Study}
\subsubsection{Efficacy of Adaptive Causal Masking (RQ2)}

We evaluated the efficacy of adaptive contextual masking by testing MaskRDT and CDT4Rec across various context lengths: [2, 4, 8, 16, 32, 64], as depicted in~\Cref{fig:images}. Our analysis reveals a pronounced divergence in performance trends between the two models as the context length increases.

CDT4Rec shows promising results at shorter context lengths but its performance declines with longer contexts. MaskRDT, on the other hand, displays remarkable consistency across the spectrum of context lengths. It matches CDT4Rec at shorter lengths and starts to show superior performance at extended context lengths, particularly from 16 onwards.

In Recommender Systems, the relevancy of historical data to current user interests varies. Shorter user interaction histories often align closely with present preferences, while older interactions may lose relevance due to the dynamic nature of user interests. CDT4Rec, with its fixed context length design, may inadvertently factor in less pertinent historical data as context lengthens, introducing noise and potentially degrading its predictive accuracy.

Contrastingly, MaskRDT is engineered to adapt to varying context lengths, allowing it to adeptly balance recent and long-term user behaviors. This flexibility could account for the observed performance dip in CDT4Rec at longer context lengths, while MaskRDT enhances performance. This is likely because the increase in context length introduces a wider range of trajectory segment lengths, enriching MaskRDT's learning with a broader diversity of user behaviors.

\begin{table}[]
\label{tab:table_cost}
\caption{Training cost of DT, CDT4Rec, and
MaskRDT. We report memory consumption and training time.}
\begin{tabular}{c|c|ccc}
\hline
\multirow{3}{*}{Model} & \multirow{3}{*}{Memory (GB)} & \multicolumn{3}{c}{Training Time (s)}                \\ \cline{3-5} 
                       &                              & \multicolumn{3}{c}{Context Length}                   \\
                       &                              & 2               & 8               & 32               \\ \hline
DT                     & 3.51                         & 1353.67         & 1722.05         & 2218.65          \\
CDT4Rec                & 3.6                          & 1685.70         & 1849.38         & 2522.59          \\
MaskRDT                & \textbf{3.1}                 & \textbf{898.85} & \textbf{905.28} & \textbf{1083.89} \\ \hline
\end{tabular}
\end{table}

\subsubsection{Assessing the Training Efficiency of MaskRDT (RQ4)}

In our investigation into the training efficiency of MaskRDT, we compared the memory consumption and training time with two other models: DT and CDT4Rec. This examination was conducted across varying context lengths to ascertain the effectiveness of our model's architecture in terms of training costs.

\Cref{tab:table_cost} presents a compelling case for the efficiency of MaskRDT. It reports lower memory usage and faster training times than its counterparts at all examined context lengths, as shown in~\Cref{tab:table_cost}. MaskRDT demonstrates the benefits of its network architecture, achieving notable training speed improvements which are crucial for practical applications where time is a valuable resource.

The results confirm that MaskRDT's training efficiency gains do not come at the expense of performance. This equilibrium of speed and accuracy is critical for models intended for real-world implementation, where both factors are essential. MaskRDT stands out as a model that can deliver swift training cycles while maintaining high-quality recommendations, making it an attractive option for scalable recommender systems.

\subsubsection{Dataset Size Influence on MaskRDT Performance (RQ4)}


In ~\Cref{fig:num_of_traje}, we explore how MaskRDT's performance is influenced by dataset size, where size is indicated by the number of user trajectories. This analysis directly addresses the question of how varying dataset sizes affect the efficacy of MaskRDT, with comparative insights drawn against DT and CDT4Rec.

The data reveals that MaskRDT maintains a high level of performance across different dataset sizes, which not only indicates stability but also reflects the model's proficiency in learning from varying context lengths. Such capability to adapt to different trajectory lengths suggests that MaskRDT is adept at understanding and incorporating both short-term and long-term user behaviors into its recommendation process. While CDT4Rec shows robust performance, it occasionally experiences fluctuations, and DT tends to stabilize only when presented with larger datasets, such as the 30k trajectory set. 


\section{Related Work}
\vspace{1mm}\noindent\textbf{RL-based Recommender Systems.}
Reinforcement learning (RL) has recently emerged as a powerful tool in the domain of recommender systems~\cite{chen2021survey}. 
\citet{zhao2018deep} introduced a RL-based page-wise recommendation framework using real-time user feedback. \citet{bai2019model} proposed a model-based technique with generative adversarial training to learn user behaviors and update recommendation policies. \citet{chen2020knowledge} integrated knowledge graphs into RL to enhance decision-making. \citet{chen2021generative} employed generative inverse reinforcement learning for online recommendations, extracting a reward function from user behaviors. 
While a significant portion of the literature has been dedicated to online RL-based recommender systems, there's a growing interest in offline RL approaches. 
\citet{chen2022off} scaled an off-policy actor-critic algorithm for industrial recommendation systems, addressing offline evaluation challenges. 
\citet{10.1145/3539618.3591636} discussed the Matthew effect in offline RL-based systems, where popular items overshadow less popular ones due to frequent recommendations. \citet{Wang2023} proposed to use causal transformers for offline RL recommender systems, addressing the challenges of reward function design and handling vast datasets.

\vspace{1mm}\noindent\textbf{Transformer in Recommender Systems.}
The transformative potential of transformer architectures has recently garnered significant attention in sequential recommendation systems. 
\citet{sun2019bert4rec} introduced BERT4Rec, leveraging a bidirectional self-attention network to capture user behavior sequences for sequential recommendations.
\citet{wu2020sse} developed a personalized transformer model, enhancing self-attentive neural architectures with SSE regularization for tailored recommendations.
\citet{chen2019behavior} employed a self-attention mechanism to enrich item representations in user behavior sequences, considering the inherent sequential patterns, and demonstrated its efficacy on a real-world e-commerce platform.
Lastly, \citet{10.1145/3543507.3583418} presented the Decision Transformer (DT) optimized for user retention, utilizing a weighted contrastive learning approach to maximize knowledge extraction from samples and prioritize high-reward recommendations.

In contrast to the aforementioned works, our proposed MaskRDT uniquely integrates adaptive causal masking with retentive networks, offering enhanced stability and performance across varying trajectory lengths in offline RL scenarios.

\section{Conclusion}
In this study, we presented the MaskRDT framework, a novel approach that integrates adaptive causal masking with retentive networks for offline RL in recommendation systems. By reinterpreting RLRS as an inference task and leveraging segmented retention mechanisms, we achieved computational efficiency and adaptability to diverse trajectory lengths. The causal mechanism further simplifies reward estimation based on user behaviors. While MaskRDT addresses many challenges in RLRS, future work could: 1) investigate optimal segment lengths specific to individual users, and 2) deepen our understanding of the causal implications of users' decisions, with the goal of refining the reward function estimation using accumulated trajectory data.

\bibliographystyle{ACM-Reference-Format}
\balance
\bibliography{sample-base}
\end{document}